\documentclass{kluwer}    % Specifies the document style.
\usepackage[dvips]{graphics}
\usepackage{graphicx,epsfig}

\def\gtsima{$\; \buildrel > \over \sim \;$}
\def\ltsima{$\; \buildrel < \over \sim \;$}
\def\prosima{$\; \buildrel \propto \over \sim \;$}
\def\gsim{\lower.5ex\hbox{\gtsima}}
\def\lsim{\lower.5ex\hbox{\ltsima}}
\def\simgt{\lower.5ex\hbox{\gtsima}}
\def\simlt{\lower.5ex\hbox{\ltsima}}
\def\simpr{\lower.5ex\hbox{\prosima}}

\begin{document}                                                                                   
\begin{article}
\begin{opening}         
\title{Is the deficit of $z>1$ field ellipticals real ?} 
\author{Emanuele \surname{Daddi}
\thanks{
The work presented here was done in collaboration with A.
Cimatti, L. Pozzetti, H. Hoekstra, H. R\"ottgering, G.
Zamorani, A. Renzini and F. Mannucci.
}
}  
\institute{Universit\'a di Firenze, Dipartimento di Astronomia e Scienze
dello Spazio}
\runningauthor{Emanuele Daddi}
\runningtitle{Is the deficit of $z>1$ ellipticals real ?}
\date{May 27, 2000}

\begin{abstract}
The results of a 700 arcmin$^2$ survey for EROs are presented. The sky
distribution of EROs is very inhomogeneous and their two point
correlation function shows a very high amplitude, a factor of 10 larger
than that of the field galaxy population. Such a clustering
can explain the strong variance found in previous works, and provides
evidence that ERO are mostly made of high--$z$ ellipticals.
The surface density of EROs found in our large survey is in good agreement 
with that expected for passively evolving ellipticals formed at high redshift
(z\simgt2.5).
\end{abstract}

\end{opening}           

%\section{Ordinary Text}  
A fundamental test of the models for the evolution of the elliptical
galaxies is the measure of their comoving density at high redshift as
compared to the local value. Existing realizations of the hierarchical
galaxy formation models predict a substantial decline of the
elliptical's number density with redshift, as they should form by
merging of lower-mass spirals at intermediate redshifts (see e.g. Baugh
et al. 1996, Kauffmann 1996). 

Passively evolving ellipticals with $1\simlt z\simlt 2$ are expected to
have very red colors, i.e. $R-K>5$--$6$, which qualify them as EROs
(Extremely Red Objects).
Thus, the search for EROs and the measure of their surface density
in deep near-IR surveys provide clues on the number density
evolution of ellipticals. 
Anyway EROs can also be strongly dust--reddened starburst galaxies (e.g. Cimatti el al. 1998). 
The density of EROs therefore provides
an upper limit to that of high--$z$ ellipticals, even if marginal indications
exist that the fraction of dusty objects among EROs may be small
(Cimatti et al. 1999).

Several groups have claimed that there is a significant deficit of 
$z>1$ evolved field  ellipticals (e.g. Zepf 1998,
Barger et al. 1999), based on the very low surface
density of EROs recovered in $K$--selected samples. 
However, others found results 
consistent with a constant comoving density, even up to $z\sim2$ 
(e.g. Benitez et al. 1999, Broadhurst \& Bowens 2000). A wide consensus on the
reality of this alleged deficit could not be reached, as very discrepant
results were obtained from works on different fields, suggesting 
that the cosmic variance could be 
strong for high--$z$ ellipticals and that results based on small area surveys 
(ranging from about 1 arcmin$^2$ of the NICMOS HDFS to $\sim60$ arcmin$^2$
of Barger et al.) are not sufficient to reach a definitive conclusion. 

\begin{figure}
\centerline{\includegraphics[height=9cm]{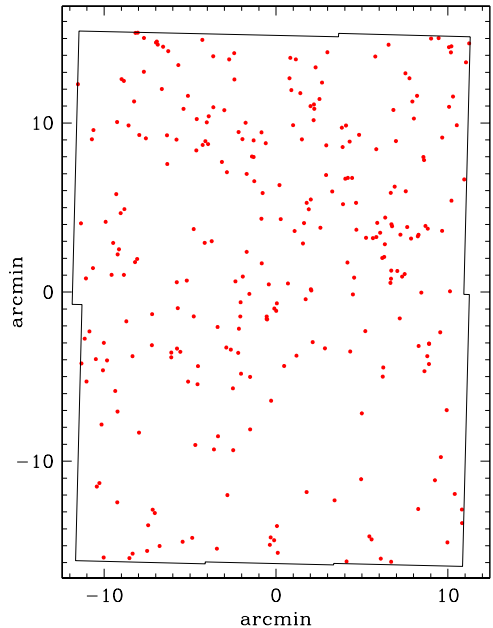}}
\caption{The sky distribution of EROs with $Ks\leq18.8$ and $R-K\geq5$.
Several tests showed that the inhomogeneities and the large {\it void}
in the bottom part of our survey are real features (see Daddi et al.
2000 for more details).}
\label{fig:1}
\end{figure}

A large area survey in the $R$ and $K$ band was therefore planned and
carried out (Daddi et al. 2000), covering about 700 arcmin$^2$ to $K\sim19$,
with the selection of a sample of $\sim400$ ($\sim50$) EROs with $R-K\geq5$
($R-K\geq6$). Fig. \ref{fig:1} shows very clearly that the
sky distribution of EROs is very inhomogeneous with overdensities and
large underdense regions, providing evidence that the
discrepant results on the ERO surface density were indeed due to the cosmic
variance.

\begin{figure}
\centerline{\includegraphics[height=8cm]{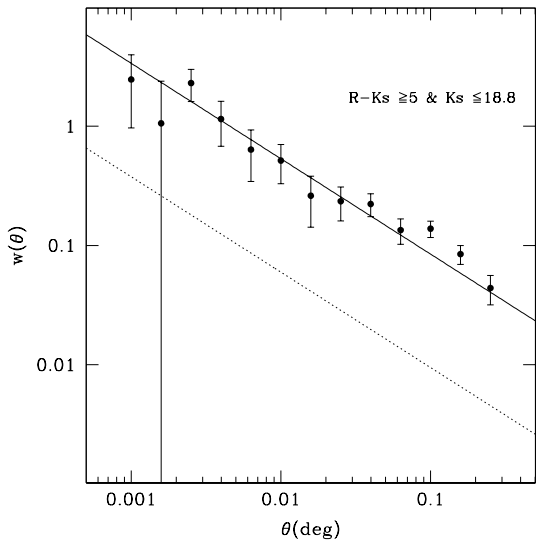}}
\caption{The measured two point correlation function for the EROs of
Fig. \ref{fig:1}. The solid line is the fitted function, with amplitude
$A(1^o)=0.014\ (\pm0.002)$ and (fixed) slope $\delta=-0.8$. The dashed line
is for the field galaxies at the same $K$ level.
}
\label{fig:2}
\end{figure}

A quantitative analysis of the two point correlation function showed
that the ERO distribution is clustered, resulting in the first
quantitative measurement of such function for the ERO population, 
and it was found to have an amplitude larger by a factor of $\sim10$ than
that of the field $K$--selected galaxies at the same $K$ magnitude limits 
(Fig. \ref{fig:2}).

The clustering of EROs provides the natural explanation of the large 
field-to-field variations of their surface density, as their
variance is increased by an additive factor proportional to
the clustering amplitude and to the square of the average number of EROs
expected. 
Our measure allows us to give a reliable estimate of the ERO variance, with
the {\it caveat} that, because of the
existence of the large underdense regions, it is much probable, on
average, to underestimate the true ERO surface density with small area
surveys. 

Even more interesting, since we expect the dusty EROs not
to have a correlation length as strong as that of ellipticals and to be
distributed on a much wider redshift range, the large--amplitude clustering 
is strong evidence that most EROs are ellipticals, 
and it allows to perform a direct comparison between the surface density
of EROs and that of the passively evolving high--$z$ ellipticals selected
with the same color and luminosity threshold (Daddi, Cimatti \&
Renzini 2000).

\begin{figure}
\centerline{\includegraphics[width=8cm]{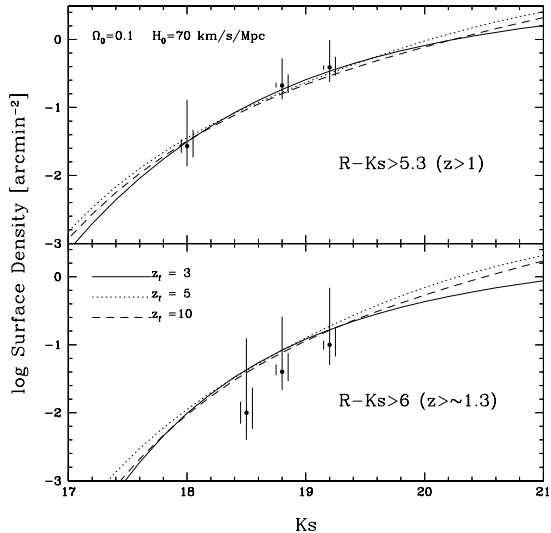}}
\caption{Comparison between the surface density of EROs measured in our
survey with the passive models predictions. The error bars are 
(from left to right) the $1\sigma$ poissonian and 
2$\sigma$ and 1$\sigma$ {\it true} 
(i.e. considering the clustering) confidence regions for the models. 
In order to estimate the variance  for the $R-K\geq6$ EROs their
clustering amplitude was derived on the basis of the color--amplitude
relation (see Daddi et al. 2000). 
}
\label{fig:3}
\end{figure}

In Fig. \ref{fig:3} such a comparison is shown for the EROs selected in our 
survey. The PLE models were computed
using the Bruzual \& Charlot 1997 synthesis models with Salpeter IMF,
solar metallicity and no dust reddening. The Marzke et al. (1994) 
pure-- elliptical LF was used and the cosmological parameters were
choosed to be $\Omega_0=0.1$, $H_0=70$ km s$^{-1}$ Mpc$^{-1}$. 
The results do not significantly change with a 
$\Omega=0.3$, $\Lambda=0.7$ cosmology.

The surface density of EROs with $R-K\geq5.3$ (expected colors of
passively evolving ellipticals with $z\simgt1$)
is fully consistent with the
passive evolution predictions, even if we allow for a reasonable
fraction of $\sim20$--30\% of EROs not to be high--$z$ ellipticals. 
Thus, we conclude that the data are
consistent with all the field ellipticals 
being in place and very red at $z\simgt1$ 
implying high formation redshift for such population ($z\simgt2.5$).
The data on the $R-K\geq6$ EROs (expected for $z\simgt1.3$) are much more 
uncertain, because of the poorer statistic, but they are still
consistent with the models at about $1\sigma$.  

%\acknowledgements

% The endnotes section will be placed here.

\theendnotes

\end{article}
\end{document}